\documentclass[10pt,a4paper]{article}
\usepackage{t1enc}
\usepackage[dvips,final]{graphicx} 
\usepackage{tabularx} 
\usepackage{amsfonts} 
\usepackage{amsthm} 

\def\du{\unskip\smash{\lower 1.4ex \hbox{\char34}}\kern-.2ex}
\def\hu{\kern-.2ex\hbox{\char92}}

\newcommand{\bdis}{\begin{displaymath}}
\newcommand{\edis}{\end{displaymath}}
\newcommand{\be}{\begin{equation}}
\newcommand{\ee}{\end{equation}}
\newcommand{\hx}{\hat{x}}
\newcommand{\pd}{\partial}
\newcommand{\hp}{\hat{p}}
\newcommand{\mbb}{\mathbb} 
\newcommand{\dd}{\rm d}

\begin{document}
\baselineskip=7.6mm
\newpage
\title{Qantum mechanics on non-commutative plane} 
\author{Michal Demetrian\footnote{{\it demetrian@fmph.uniba.sk}}
, Denis Kochan\footnote{{\it kochan@fmph.uniba.sk}} \qquad \\
{\it Department of Theoretical Physics} \\ 
{\it Faculty of Mathematics, Physics and Informatics} \\
{\it Comenius University, Bratislava, Mlynsk\' a dolina 842 48, Slovakia}}
\maketitle

\abstract{One of the simplest example of non-commutative (NC) spaces is the 
NC plane. In this article we investigate the consequences of the 
non-commutativity to the quantum mechanics on a plane. 
We derive corrections to 
the standard (commutative) Hamiltonian spectrum for hydrogen-like atom and
isotropic linear harmonic oscillator (LHO) and formulate the 
problem of the potential scattering on the NC plane.  
In the case of LHO we  
consider the 
non-commutativity of the momentum operators, too.} 
\\ 
\begin{center} PACS numbers: 03.65.G; 03.65.N; 02.40.G. \end{center} 

\section{Introduction} 
In recent years, the method of non-commutative geometry (NCG) was developed
\cite{a1} 
 and applied to various physical situations \cite{a2}. By the
results of string theory arguments \cite{sw} the non-commutative plane 
has been studied extensively. \\ 
\\ 
Two dimensional 
non-commutative quantum mechanics (NCQM) is based on a simple modification 
of  
commutation relations between the self-adjoint 
position operators ($\hat{\vec{x}}$) 
and the self-adjoint momentum 
operators ($\hat{\vec{p}}$) which satisfy 
\be \label{1} 
\begin{array}{ccc} 
\left[ \hx_a,\hx_b\right] & = & i\theta_{ab} \quad , \\ 
\left[ \hp_a,\hp_b\right] & = & 0 \quad , \\
\left[ \hx_a,\hp_b\right] & = & i\hbar \delta_{ab} \quad , 
\end{array} 
\qquad a,b,\dots \in \{ 1,2\} \quad , 
\ee
where $\theta_{ab}$ is real and 
antisymmetric, i.e. $\theta_{ab}=\theta \epsilon_{ab}$ (
$\epsilon_{ab}$ is the completely antisymmetric tensor with 
$\epsilon_{12}=1$). The spatial non-commutative 
parameter $\theta$ is of dimension of $(length)^2$, so $\sqrt{\theta}$ may 
be considered as the fundamental length (Planck length?).   
If $\theta$ goes to zero we obtain the standard Heisenberg algebra commutation 
relations. \\ 
\\ 
Suitable realization of commutation 
relations between the position operators (the first of eqs. (1)) 
is given  by the $\star$-product (Moyal product) \cite{mo}
defined as follows 
\be 
(f\star g)(x)=\exp\left[ \frac{i}{2}\theta_{ab}\pd_{x_a}\pd_{y_b}\right] 
f(x)g(y)|_{x=y} \quad . 
\ee 
The wave functions are taken as $\psi(\vec{x})$ and the operators
$\hat{\vec{x}}$ and $\hat{\vec{p}}$ are realized as follows 
\bdis 
\hx_a \psi(\vec{x})=x_a\star \psi(\vec{x}) \quad , \qquad
\hp_a\psi(\vec{x})=-i\hbar\pd_a\psi(\vec{x}) \quad . 
\edis
Let us remark that $|\psi(\vec{x})|^2$ can not be interpreted as the 
probability density to find the system in the configuration $(x_1,x_2)$. 
That follows from the first relation in the eq. (\ref{1}). \\ 
The quantization procedure is the same as in the standard quantum mechanics.
We replace the classical observables - the functions 
on the phase space - $A(\vec{p},\vec{x})$ 
by the 
self-adjoint operators $\hat{A}=A(\hat{\vec{p}},\hat{\vec{x}})$ 
which act on suitable Hilbert 
space. The ordering problem is like that in the standard 
quantum theory. 
The Hilbert space can be conistently taken to be the same as 
the Hilbert space of the corresponding commuting system, for example 
$L^2(\mbb{R}^2)$: squared 
integrable functions on the plane with the standard Lebesgue measure. 
The time evolution is given by the Schr\" odinger equation 
\bdis i\hbar\pd_t|\psi\rangle=\hat{H}|\psi\rangle \quad ,  \edis
where $\hat{H}=H(\hat{\vec{p}},\hat{\vec{x}})$ is the Hamiltonian. 
The only nontrivial part of such a formulation
is to give the Hamiltonian. In what follows we shall consider two dimensional 
hydrogen-like atom, isotropic linear harmonic oscillator and potential 
scattering.

\newcommand{\tx}{\tilde{x}}
\newcommand{\tp}{\tilde{p}}
\newcommand{\ep}{\epsilon}

\section{Hydrogen-like atom} 
Two dimensional hydrogen-like atom in NCQM is defined by the following
Hamiltonian (the 
Einsten's summation convention in latin indeces is used.) 
\be \label{ham} 
\hat{H}=\frac{1}{2}\hp_a\hp_a+U_0\ln\left(\frac{\sqrt{\hx_a\hx_a}}{r_0}\right)
\equiv \frac{1}{2}\hp_a\hp_a+U(\hat{\vec{x}}) \quad , 
\ee
where $U_0$ and $r_0$  are the positive constants. We define the new
operators $\tx_a$, $\tp_a$
\be 
\begin{array}{ccc} \label{2} 
\tx_a &= & \hx_a+\frac{1}{2\hbar}\theta_{ab}\hp_b \quad , \\
\tp_a & = & \hp_a \quad , 
\end{array} 
\ee
which satisfy the usual canonical commutation relations 
\be 
\begin{array}{ccc} 
\left[ \tx_a,\tx_b\right] & = & 0 \quad , \\
\left[ \tp_a,\tp_b\right] & = & 0 \quad , \\
\left[ \tx_a,\tp_b\right] & = & i\hbar\delta_{ab} \quad . 
\end{array}
\ee
The representation of the above commutation relations in the $L^2(\mbb{R}^2)$ 
is well known: 
\be 
(\tx_af)(\vec{x})=x_af(\vec{x})\quad,  \qquad
(\tp_af)(\vec{x})=-i\hbar\pd_af(\vec{x}) \quad . 
\ee
If we replace hat operators by tilde operators in the Hamiltonian (3) 
(to simplify calculations 
we shall use the system of units in which the
mass of electron and Planck constant are equal to one) 
and 
expand the potential to the Taylor series,  
we obtain the first $\theta$ - order time independent 
Schr\" odinger equation in polar coordinates $(r, \phi)$  
\be 
\Delta\psi=2\left[ U_0\ln\left( \frac{r}{r_0}\right) -\ep
+\theta\frac{iU_0}{2r^2}\pd_{\phi}\right] \psi \quad ,  
\ee
where $\ep$ is the Hamiltonian's eigenvalue. 
We note that $-i\pd_\phi$ is the 
$z$ component ($L_z$) of the angular momentum operator. Standard  
separation of variables $\psi(r,\phi)=R(r)\exp(im\phi)$ 
leads to the radial Schr\" odinger equation
\be 
\frac{1}{r}(rR^\prime)^\prime+ 
\left[ 2(\ep-U(r))-\frac{m^2-mU_0\theta}{r^2}\right] R=0 \quad , 
\ee
where $m$ (orbital quantum number) is an arbitrary integer. \\
In the case of $\theta=0$ we obtain radial equation for two-dimensional  
hydrogen-like
atom in commutative quantum mechanics. It is an important fact 
that the structure of equation
(8) does not change. So we can state that if 
\be 
\ep^C=\ep^C(n,m^2) \quad , 
\ee
is the spectrum of Hamiltonian in the commutative case then in NC quantum 
mechanics the
spectrum is given by 
\be 
\ep(n,m^2)=\ep^C(n,m^2-mU_0\theta) \quad , 
\ee
where $n$ is the principial quantum number. We have derived the approximative
formulae for the commutative spectrum (see Appendix). The results are \\
(i.) For the states with $|m|\gg n$, $n=1,2,\dots$  
\be \label{spect} 
\ep(n,m^2)=\frac{U_0}{2}\left\{ 
1+\ln\left[ \frac{m^2-mU_0\theta-1/4}{r_0^2U_0}\right] +\sqrt{2}\frac{n-1/2}
{\sqrt{m^2-mU_0\theta-1/4}}\right\} \quad . 
\ee
(ii.) For high excited ($n\gg 1$) states with zero orbital momentum ($m=0$) 
\be 
\ep(n,m^2=0)=\frac{U_0}{2}\left\{ 
\ln\left[ \frac{2\pi}{U_0r_0^2}\right] +2\ln(n-1/2)\right\} \quad . 
\ee

\subsection*{Linear Stark effect} 
In the commutative theory 
the potential energy of an electron in an external electrostatic field 
oriented along the $x_1$ axis is given by $\delta\hat{H}=eEx_1$, where 
$e$ is the electric charge of 
the electron and $E$ is the intensity of the electric 
field. In the non-commutative theory it holds  
\be \label{st} 
\delta\hat{H}=eE\hx_1=eE(x_1+i\frac{\theta}{2}\partial_{x_2}) \quad . 
\ee 
$\delta\hat{H}$ is considered as the small perturbation to the Hamiltonian 
(\ref{ham}). Linear Stark effect is the change in the hydrogen-like atom 
energy levels due to the perturbation 
shift of (\ref{st}) computed within the 
first order of the perturbation theory. It contains two contributions, the 
commutative one: $\delta\epsilon^C_{nm}=\langle nm|eEx|nm\rangle$ and the 
noncommutative one: $\delta\epsilon^{NC}_{nm}=\langle 
nm|eEi\frac{\theta}{2}\partial_{x_2}|nm\rangle$. They both are equal to 
zero. It follows from the fact that the eigenstates $|nm\rangle$ of the 
unperturbated Hamiltonian (\ref{ham}) are not degenerated - this fact 
allows us to write the energetical changes as we have done above.  
Indeed, the 
states $|nm\rangle$ are of the form $f_{nm}(r)e^{im\phi}$, so we have 
\bdis \delta\epsilon^C_{nm}\sim \int_{-\infty}^{+\infty}\dd x_1
\int_{-\infty}^{+\infty}\dd x_2 f_{nm}^*e^{-im\phi}x_1f_{nm}e^{im\phi}=0 
\quad , 
\edis 
and 
\bdis \delta\epsilon^{NC}_{nm}\sim \int_{-\infty}^{+\infty}\dd x_1
\int_{-\infty}^{+\infty}\dd x_2 f_{nm}^*e^{-im\phi}
\left(\frac{x_2}{r}f'_{nm}e^{im\phi}+f_{nm}ime^{im\phi}\frac{x_1}{r^2}\right)
=0 \quad . 
\edis

\newcommand{\si}{\sigma}
\newcommand{\om}{\omega} 
\newcommand{\kp}{\kappa} 

\section{Linear harmonic oscillator} 
Now, 
let us consider a slight generalization of the commutation relations (1) 
between the canonically conjugated positions and momenta 
\be \label{13} 
\begin{array}{ccc}
\left[ \hx_a,\hx_b\right] & = & i\theta_{ab} \quad , \\
\left[ \hp_a,\hp_b\right] & = & i\kappa_{ab} \quad , \\
\left[ \hx_a,\hp_b\right] & = & i\hbar
\left( 1+\frac{\theta\kappa}{4\hbar^2}\right) \delta_{ab} \quad . 
\end{array}
\ee 
The role of 
$\kappa_{ab}=\kappa\ep_{ab}$ is similar that of $\theta_{ab}$ with the 
difference between the two quantities 
that $\kappa$ is of dimension of $(momentum)^2$ and 
$\sqrt{\kappa}$ plays the role of fundamental momentum. We stress that  
the eq. (\ref{13}) with $\theta_{ab}=0$ is valid for 
a particle moving in a magnetic field in the standard quantum mechanics, 
but in this case the momenta 
are not conjugated to coordinates as it is the case here. \\ 
We would like to find some linear transformation 
$(\tilde{\vec{x}},\tilde{\vec{p}})\rightarrow
(\hat{\vec{x}},\hat{\vec{p}})$ so that tilde operators satisfy canonical
commutation relations (5). In other words, we would like to transform the 
canonical 
form defined by (5) to the form defined by (\ref{13}). 
It is possible 
only if $4\hbar^2\not= \kappa\theta$, because in this case the form defined 
by (\ref{13}) is singular. 
The transformation ($T$) in question is of the form 
\be  \label{14}  
\begin{array}{ccc}
\hx_a & = & \tx_a-\frac{\theta}{2\hbar}\epsilon_{ab}\tp_b \quad , \\ 
\hp_a & = & \tp_a+\frac{\kappa}{2\hbar}\epsilon_{ab}\tx_b   \quad . 
\end{array} 
\ee 
Any other transformation $T'$ 
of the tilde operators to the hat operators can 
be written as $T'=T\circ U$, where $U$ is from the group of transformations 
which preserve the form defined by (5), i.e. $U\in Sp(2,\mathbb{R})$ (for 
more informations about the symplectic groups and their use in physics 
see for example \cite{wood}). 
The operator realization of tilde operators in $L^2(\mathbb{R}^2)$ 
is the standard one given by (6). \\  
Let us consider the isotropic linear harmonic oscillator Hamiltonian in the
NC quantum mechanics  
\be 
\hat{H}=\frac{1}{2M}\hp_a\hp_a+\frac{M}{2}\om^2\hx_a\hx_a \quad . 
\ee
We express the hat operators in the LHO Hamiltonian in terms of the tilde
operators using the formulae (\ref{14}). We get 
\begin{eqnarray}  
\hat{H} & = & \tp_a\tp_a
\left( \frac{1}{2M}+\frac{1}{2}M\omega^2\frac{\theta^2}{4\hbar^2}\right) + 
\tx_a\tx_a
\left( \frac{1}{2}M\omega^2+\frac{1}{2M}\frac{\kappa^2}{4\hbar^2}\right)  
\\ \nonumber & & 
+\tp_a\epsilon_{ab}\tx_b
\left( 
\frac{1}{2M}\frac{\kp}{\hbar}+\frac{1}{2}M\om^2\frac{\theta}{\hbar}\right) 
\quad . 
\end{eqnarray} 
The last term in the Hamiltonian is proportional to the $z$ component of the
angular momentum ($L_z=\ep_{ab}\tx_a\tp_p$). The spectrum $\ep_{n_1n_2}$ 
of this type of Hamiltonian is well known, so
we shall write down the result 
\begin{eqnarray} \label{lospec}  
\epsilon_{n_1n_2} & = & \sqrt{\hbar^2\om^2
\left( 1+\frac{\kappa\theta}{4\hbar^2}\right)^2+\frac{1}{4}
\left( \frac{\kappa}{M}-\theta M\om^2\right)^2}(n_1+n_2+1) \\ \nonumber 
 & & -\frac{1}{2}
\left( \frac{\kappa}{M}+\theta M\om^2\right)^2 (n_1-n_2) \quad , 
\end{eqnarray} 
where $n_1,n_2=0,1,2\dots $. \\ 
If $\kappa$ and $\theta$ aprroach zero  we
recover the standard spectrum for the two-dimensional LHO.

\newcommand{\vr}{\vec{r}}

\section{Potential scattering on NC - plane} 

The theory of the best-known quantum systems - linear harmonic 
oscillator and hydrogen-like atom is formulated 
in the two previous sections of this 
work. It is natural that the next step will be the formulation of the 
potential scattering on non-commutative plane. Our goal is to give the NC 
correction to the commutative cross-section. \\ 
In the first step we remind the computation of the differential 
cross-section in two-dimensional 
commutative quantum mechanics. Let $V(\vr), \quad 
\vr\in \mathbb{R}^2$ be the potential energy vanishing at infinity. 
We consider the particle of the mass $M$ incident (coming  from the 
infinity) with the wave vector $\vec{k}$, say $\vec{k}=(k,0)$. We expect 
the out-state (at $r\to \infty$) will be of the form 
\be \label{as} 
\psi(\vr)\sim e^{ikx}+f(\phi)\frac{e^{ikr}}{\sqrt{r}} \quad , 
\ee  
where $r=|\vr |$ and a function $f$ depends on a 
variable $\phi\in [0,2\pi )$ - the polar angle. The differential 
cross-section $\dd \sigma/\dd \phi$ is expressed in terms of $f$ as follows 
\be 
\frac{\dd \sigma}{\dd \phi}=|f(\phi )|^2 \quad . 
\ee
We have to solve the following integral equation  
\be \label{ie} 
\psi(\vr)=e^{ikx}+\int \dd ^2\vr' G_k(\vr,\vr')U(\vr')\psi(\vr') \quad , 
\ee 
where $U=\frac{2M}{\hbar^2}V$ and 
$G_k$ is the Green's function of the operator $\Delta^{(2)}+k^2$ 
obeying the boundary condition (\ref{as}) \footnote{$G$ is given by the 
formula $G(\vr,\vr')=G(|\vr-\vr'|)=\frac{1}{4i}H^{(1)}_0(k|\vr-\vr'|)$, 
where $H_0^{(1)}$ is the Hankel's function.}. The solution to the eq. 
(\ref{ie}) in the Born approximation is given by 
\be   
\psi(\vr)=e^{ikx}+\frac{e^{ikr}}{\sqrt{r}}\frac{e^{-i\pi/4}}
{i\sqrt{8\pi k}}
\int \dd ^2\vr' 
e^{-i\vec{k}'.\vr'}U(\vr')e^{i\vec{k}.\vr'} 
 \quad ,  
\ee 
where $\vec{k}'=k\frac{\vr}{r} \quad \mbox{and} \quad \vec{k}=(k,0)$. 
We identify the function $f$ from the above equation and the eq. 
(\ref{as}), and we get 
\be  \label{fifi} 
f(\phi)=\frac{e^{-i\pi/4}}{i\sqrt{8\pi k}}\int \dd ^2\vr' 
e^{-i\vec{k}'\vr'}U(\vr')e^{i\vec{k}\vr'}=
\frac{e^{-i\pi/4}}{i\sqrt{8\pi k}}\int \dd ^2\vr' 
e^{i\vec{q}.\vr'}U(\vr') \quad ,    
\ee
where $\vec{q}=\vec{k}-\vec{k}'$. The right-hand side of the eq. 
(\ref{fifi}) depends on the scattering angle $\phi$ via $q$, because 
$q=2k\sin(\phi/2)$. \\ 
In the case of a radial symetric function $V$ we have 
\begin{eqnarray} 
f(\phi) & = & \frac{e^{-i\pi/4}}{i\sqrt{8\pi k}}
\int_0^\infty \dd r'r'U(r')\int_0^{2\pi}\dd \alpha  
e^{iqr'\cos(\alpha)} = \\ \nonumber & & 
\frac{2\pi e^{-i\pi/4}}{i\sqrt{8\pi k}}\int_0^\infty \dd r'r'U(r')J_0(qr') 
 \quad , 
\end{eqnarray}
where $J_0$ is the Bessel's function of the first kind\footnote{We shall 
use this Bessel's 
function in what follows, so we remind that for any complex $x$ the value  
of $J_0(x)$ is 
\bdis J_0(x)=\sum_{k=0}^{\infty}\frac{(-1)^k}{(k!)^2}\left( 
\frac{x}{2}\right)^2=\frac{1}{2\pi}\int_0^{2\pi}\dd \alpha 
e^{ix\cos(\alpha)} \quad . \edis }. \\  
The noncommutativity of the plane according to the eqs. (\ref{1}) can be 
implemented to the scattering problem in the same way 
as it has been done  
in the introduction and the second section. We start from the Schr\" 
odinger equation $H\psi=E\psi$, where 
$H=\hat{p}_a\hat{p}_a/(2M)+V(\hat{\vec{r}})=
\tp_a\tp_a/(2M)+V(\hat{\vec{r}})$. Then we replace the hat operators in 
$V$ by the tilde operators according to the transformations (\ref{2}) and we 
perform the expansion of the function $V$ into the Taylor series in powers 
of $\theta$. In the special case of the radial symetric potential $V$ we 
have 
\be 
V(\sqrt{\hx_a\hx_a})= 
V(r)-\frac{\theta}{2\hbar}\frac{1}{r}\frac{\dd V(r)}
{\dd r}\epsilon_{ab}\tx_a\tp_
b +O(\theta^2) 
 \quad . 
\ee 
So, we can state, that the first NC correction to the $f$ function is given 
by the following formula 
\be 
f^{NC}(\phi)-f(\phi)=\frac{\theta}{2}\frac{e^{-i\pi/4}}{\sqrt{8\pi 
k}}\int \dd ^2\vr' 
e^{i\vec{q}.\vr'}\frac{1}{r'}\frac{\dd U(r')}
{\dd r'}\epsilon_{ab}x'_ak_b \quad 
. 
\ee 
After some rearrargements we get the final formula 
\be \label{scattam} 
f^{NC}(\phi)-f(\phi)=-\frac{\theta}{2}\frac{i\pi 
e^{-i\pi/4}}{\sqrt{8\pi}}\frac{1}{\sqrt{k}}\cot(\frac{\phi}{2})\int_0^
\infty \dd r' r' \frac{\dd U(r')}{\dd r'}
\frac{\dd J_0(qr')}{\dd r'} \quad . 
\ee 

Note that for the potential proportional to $\ln(r)$ the scattering 
states do not exist, so there is no analogue of the Rutherford formula 
within the two-dimensional 
quantum mechanics. But one can investigate the 
scattering of the charged particle (electron) 
on the neutral atom and obtain the NC correrction.

\section{Conclusion} 
In this paper we have presented the results on the two-dimensional 
systems within 
non-commutative quantum mechanics for the hydrogen-like atom, 
the isotropic LHO and the potential scattering problem. 
We have obtained the corrections depending on the parameter of the space 
noncommutativity  
to the classical  
(commutative) spectra of related Hamiltonians and cross-section. 
We note that two-dimensional hydrogen-like 
atom has the analogy in three-dimensional 
motion of a charged particle around the 
homogenous charged straight line. Some others interesting 
results in NC quantum 
theory can be found in \cite{sh},\cite{9}-\cite{14}. \\ 
\\ 
The main goal of the NC quantum mechanics 
is to find a measurable effect. Unfortunately, we
cannot hope that our corrections to the energy levels of the systems in
question could be experimentally verified because of its dimensionality (LHO)
and experimental complications with the realization of the 
three-dimensional 
analogy of 
the hydrogen-like atom. In \cite{sh} the modification of the 
energy levels and Lamb 
shift for three-dimensional 
hydrogen-like atom due to the presence of the non-commutative  
plane in $\mathbb{R}^3$ 
were presented as "measurable". But, this type of space 
non-commutativity is not applicable if we insist on the isotropy 
of the space. 
In spite of this it would be desirable to use three-dimensional space 
non-commutativity which 
preserves rotational symmetry of the hydrogen-like atom problem. \\ 
\\ 
The starting point of our treatment to the NC quantum mechanics 
are the nontrivial 
commutation relations between the position operators. The violation of the 
space parity is explicitly shown in the formulae (\ref{spect}), 
(\ref{lospec}) and (\ref{scattam}). It is well-known 
that the parity is 
violated if one considers the external magnetic field as well as 
that the CP is preserved in this case. In our two noncommutative cases 
(commutative/noncommutative momenta) the space parity as well as the CP are 
violated. The reason of the CP violation is that $\theta$ (and 
$\kappa$ too) are not related to the electric charge as 
the magnetic field. 
In addition, we have found an analogue to the relativistic 
quantum mechanics in the (perturbative) formula 
(\ref{spect}): one obtains nonzero imaginary part of the energy 
for very large values of $U_0\theta$ and properly small $m$. 
This effect is well-known from the theories of Klein-Gordon
and Dirac hydrogen-like atom. \\ 
{\bf Acknowledgement.} 
We are grateful to P. Pre\v snajder and 
P. \v Severa for useful discussions. We
are especially thankful to V. Balek for his help with two-dimensional 
hydrogen-like atom.

\newcommand{\Om}{\Omega}
\newcommand{\hr}{\hat{R}}

\section*{Appendix}     
In this section we briefly describe how to find the formulae (11) and (12).
Let us start with the equation (8), whose structure does not change 
when we put 
$\theta=0$ (the commutative case). \\
Let us first analyze the case of $m\not= 0$. If we introduce into eq.
(8) a new function $\chi(r)=\sqrt{r}R(r)$, we get  
\be 
\chi^{\prime\prime}+\left[ 2\ep-2U_{eff}\right]\chi=0 \quad , 
\qquad \chi\in L^2(R^+)
\quad , 
\ee
where the effective potential $U_{eff}$ is given by 
\bdis 
U_{eff}=U+\frac{m^2-1/4}{2r^2} \quad . 
\edis 
We expand $U_{eff}$ into the powers of $(r-r_k)$,  where
$r_k=[(m^2-1/4)/U_0]^{1/2}$  
is the point at 
which $U_{eff}$ is minimal. Harmonic approximation 
gives 
\be 
U_{eff}^{h}(r)=U_{eff}(r_k)+\frac{1}{2}\Om^2(r-r_k)^2 \quad , \qquad
\Om^2=2\frac{U_0^2}{m^2-1/4} \quad . 
\ee
It can be easily shown that the harmonic approximation is valid  
only if the following inequality is fulfilled
\be 
\mid \frac{U_{eff}-U_{eff}^h}{U_{eff}^h}~\mid \approx 
\frac{5}{3}\frac{(2n+1)^{1/2}}{(2m^2-1/2)^{1/4}}\ll 1 \quad ,  
\ee
where $n=1,2,\dots$ is the principal quantum number. For the LHO with the
effective potential (27) one obtains eq.(11) with $\theta=0$. \\
In the second step we analyze the high excited states with zero angular
momentum $m$. Substitution 
\bdis 
v=\ln\left( \frac{r}{r_0}\right)-\frac{\ep}{U_0} \quad , 
\qquad v\in R \quad , 
\edis 
in (8) leads to 
\be 
\hr^{\prime\prime}(v)-\left[ \Xi v e^{2v}\right] \hr(v)=0 \quad , \quad
\Xi(\ep)=2U_0r_0^2\exp\left( \frac{2\ep}{U_0}\right) \quad . 
\ee 
Now, we shall investigate the asymtotic solutions to the above boundary
problem. In the case of $v\to -\infty$ the only acceptable solution is 
\be 
\hr(v)_-=const \quad . 
\ee 
In the other extreme case of $v\to +\infty$ we have the WKB \cite{LL} 
solution 
\be 
\hr(v)_+\sim V_0^{-1/4}\exp\left( -V_0^{1/2}\right) \quad , 
\qquad V_0=\Xi ve^{2v} \quad . 
\ee 
It is important that the WKB limit is applicable  
for $v>v_m$, where $v_m$ is determined by the standard WKB condition 
\be 
\frac{1}{2V_0}\mid \frac{\dd (V_0)^{1/2}}{\dd v}~\mid\ll 1 \quad . 
\ee
For high excited states with $\Xi \gg 1$  
$v_m$ can be approximated by 
\be 
|v_m|\sim \ln(\Xi) \qquad \mbox{and} \qquad v_m<0 \quad . 
\ee 
The full WKB function $\hr^{WKB}$ is then of the form 
\be 
\hr(v)^{WKB}= \left\{ \begin{array}{ccc} 
C_0p^{-1/2}\sin\left( \int_v^0 p^\prime \dd v^\prime 
+\pi/4\right) & ; & v_m<v<0 \quad , \\ 
 & & \\
\frac{1}{2}C_0q^{-1/2}\exp\left( -\int_0^vq^\prime \dd v^\prime\right) 
& ; & v>0
\quad , 
\end{array} \right. 
\ee
where $p=(-V_0)^{1/2}$ and $q=(V_0)^{1/2}$. The continuity condition of the
wave function in $v_m$ leads to the following analogy of the
Bohr-Sommerfeld quantization rule 
\be 
\sqrt{\Xi}\int_{-\infty}^0|v|^{1/2}e^v\dd v=\pi(n-1/2) \quad , 
\ee
from which we, using (29), 
obtain instantly the spectrum formula (12).

\end{document}